\documentclass[conference]{IEEEtran}
\IEEEoverridecommandlockouts

\usepackage{adjustbox}
\usepackage{dashbox}
\usepackage{colortbl}
\usepackage{booktabs} 
\usepackage{makecell}
\usepackage{cite}
\usepackage{enumitem}
\usepackage{amsmath,amssymb,amsfonts}
\usepackage{cite}
\usepackage{graphicx}
\usepackage{subfig}
\usepackage{caption}
\usepackage{textcomp}
\usepackage[linesnumbered,ruled]{algorithm2e}
\def\BibTeX{{\rm B\kern-.05em{\sc i\kern-.025em b}\kern-.08em
    T\kern-.1667em\lower.7ex\hbox{E}\kern-.125emX}}

\makeatletter
\def\endthebibliography{%
  \def\@noitemerr{\@latex@warning{Empty `thebibliography' environment}}%
  \endlist
}
\makeatother
    
\begin{document}

\title{On Event Causality Detection in Tweets}

\author{\IEEEauthorblockN{Humayun Kayesh}
\IEEEauthorblockA{\textit{School of ICT} \\
\textit{Griffith University, Australia}\\
humayun.kayesh@griffithuni.edu.au}
\and
\IEEEauthorblockN{Md. Saiful Islam}
\IEEEauthorblockA{\textit{School of ICT} \\
\textit{Griffith University, Australia}\\
saiful.islam@griffith.edu.au}
\and
\IEEEauthorblockN{Junhu Wang}
\IEEEauthorblockA{\textit{School of ICT} \\
\textit{Griffith University, Australia}\\
j.wang@griffith.edu.au}

}

\maketitle

\begin{abstract}
Nowadays, Twitter has become a great source of user-generated information about events. Very often people report causal relationships between events in their tweets. Automatic detection of causality information in these events might play an important role in predictive event analytics. Existing approaches include both rule-based and data-driven supervised methods. However, it is challenging to correctly identify event causality using only linguistic rules due to the highly unstructured nature and grammatical incorrectness of social media short text such as tweets. Also, it is difficult to develop a data-driven supervised method for event causality detection in tweets due to insufficient contextual information. This paper proposes a novel event context word extension technique based on background knowledge. To demonstrate the effectiveness of our proposed event context word extension technique, we develop a feed-forward neural network based approach to detect event causality from tweets. Extensive experiments demonstrate the superiority of our approach.

\end{abstract}


\section{Intruduction}
  
\begin{figure}
        \centering
        \includegraphics[scale=0.55]{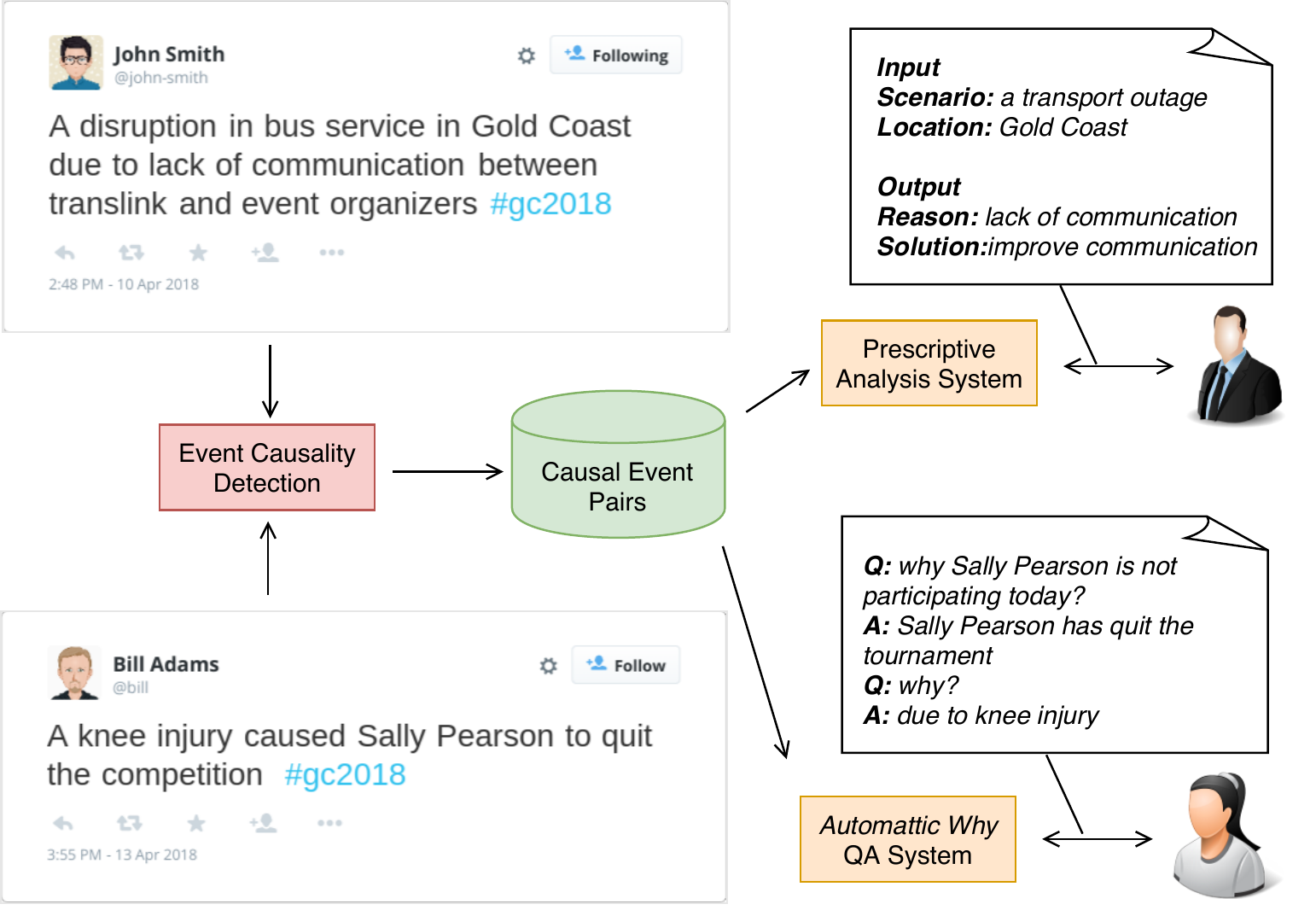}
        \vspace{-1ex}
        \caption{Application of automatic event causality detection in \emph{why} question answering and predictive event analysis}        
        \label{fig:app-scenerios}
        \vspace{-3ex}
    \end{figure}

Microblogging sites such as Twitter has become a popular medium for users to express their opinion and respond to different situations. Therefore, tweets can be an important source of causality information between events and this information might play an important role in predictive event analytics. For example, the following tweet ``\textit{\textbf{A disruption in bus service in Gold Coast} due to \textbf{lack of communication between translink and event organizers}}" contains two causally related events. From this tweet it can be said that the ``lack of communication" was a cause of transport service disruption in Gold Coast. This information could be applied in prescriptive analytics by the decision makers to reduce the chance of a future transport disruption during public gatherings. Causality information can also be applied to improve automated \textit{why} question answering. For example, we can answer the question ``Why Sally Pearson is not participating today?" from the event causality information extracted from the following tweet, ``\textit{\textbf{A knee injury} caused \textbf{Sally Pearson to quit the competition}}". The uses of the above event causalities in predictive event analytics based applications are visualized in Fig. \ref{fig:app-scenerios}. 
   
The extraction of causal relationships is an evolving area of research \cite{Blanco2008, Do2011, Ciresan2012, Riaz2013, Riaz2014, Mirza2014, Luo2016, Kruengkrai2017}. Existing approaches often apply linguistic rules or commonsense knowledge to identify causal relationship from short text. However, processing tweets is more challenging than normal short text \cite{Ritter2011}. One of the most obvious challenges is that tweets are unstructured and highly informal in nature. Hence, the linguistic rule-based approaches \cite{Luo2016, Sasaki2017}, which depend on grammatical correctness of text, perform poorly on tweets (see section \ref{sec:experiments}). Additionally, the causally related event pairs appear infrequently in tweets and the causal relationship between events lacks context information. Due to this lack of adequate context information, supervised learning based approaches such as \cite{Ponti2017} are not much effective on tweets. In this paper, we propose an automated system to detect causally related event pairs from social media short text, e.g. tweets. In our proposal, we use a feed-forward neural network to detect causal relationship between events to deal with the unstructured nature of the data. To accurately train the model, we propose a context word extension method to enhance the feature set. We extend the event keywords of both candidate causal event and candidate effect event using background knowledge before applying the model. The background knowledge is captured by creating a causal network from news article text using a set of causal cue words. To be specific, the main contributions of this paper are as follows:

\begin{enumerate}

\item we propose a novel event context word extension technique that uses commonsense background knowledge extracted from news articles to enhance the feature set; 

\item we develop a neural network based event causality detection method to detect causality relationship between events; 

\item we perform an extensive experimental evaluation to demonstrate that our proposed event context word extension technique outperforms the method of using only the event keyword and other attribute words; and

\item we also compare the performance of our method with existing methods of causality detection from short text.
\end{enumerate}

The rest of the paper is organized as follows: Section \ref{sec:relatedwork} discuses the related work; Section \ref{sec:problem-formulation} formulates the problem studied in this paper; Section \ref{sec:proposed-metnod} illustrates the proposed method; Section \ref{sec:experiments} discusses the experimental results and finally, Section \ref{sec:conclusion} concludes the paper.

\section{Related Work}
\label{sec:relatedwork}
  
Causality detection approaches can be categorized into two broad categories: extracting causality between phrases and event-based causality detection. One of the early approaches that identifies explicit, implicit and non-causal relationships between verb-verb pairs is proposed in \cite{Riaz2013}. The authors propose a technique to automatically generate training corpus of causal relationship between verbs. A supervised model is trained on the corpus to predict causal strength between verbs in a pair. Later, in \cite{Riaz2014} the authors extend the previous approach and extract causality relationship between noun-verb pairs. The authors at first identify all the nouns and verbs from a sentence and then apply a supervised classifier to identify causality between grammatically connected noun-verb pairs. For training, the authors use lexical, semantic and structural features. A more recent approach \cite{Luo2016} builds a causality network of terms from a collection of web text. The authors apply linguistic rules, e.g., `A \textit{causes} B' to extract the terms with causal relations. They create a directed graph where each node represents a term, and each edge contains causal co-occurrence score. Finally, the co-occurrence scores between terms in the causal graph are used to compute the causal strength. The final causal relationship is calculated by using the co-occurrence score. This approach is extended by Sasaki et al. \cite{Sasaki2017} for multi-word terms where authors calculated causal strength not only for the pairs of single words but also for the multi-word pairs. Although the above approaches calculate causal strengths between phrases, they do not take events into account.

On the other hand, some approaches extract causal relationship betweenvent pairs. \cite{Do2011} proposes a method that uses distributional probability and discourse connectives to detect event causality. The authors represent events as $p$($a_1$, $a_2$, $a_3$,..., $a_m$), where $p$ denotes the predicate or the trigger word and $a_i$ represents the event attribute. A trigger word can be a \emph{noun} or a \emph{verb}. The authors calculate \emph{cause-effect association} score between a pair of events by calculating causal associations for predicate-predicate, predicate-argument and argument-argument. To calculate predicate-predicate association, the authors use \emph{pointwise mutual information} (PMI) \cite{Chambers2008} score along with \emph{inverse document frequency} (IDF), syntactical distance between predicates and co-occurrence probability. The average PMI score between each pair of terms is calculated to estimate the predicate-argument association. Similarly, the argument-argument association is calculated by multiplying the PMI value for each pair and then, dividing it by the multiplication of number of arguments in the two events. Additionally, the authors identify discourse relation in text using \emph{penn discourse treebank} (PDTB) \cite{Prasad2007}. However, the PMI based approaches are sensitive to co-occurance frequency and do not perform well for infrequent events\cite{turney2010}. Another approach proposed by Mirza \cite{Mirza2014} extracts temporal and causal relationships between events, and assumes that cause must precede the effect and propose a methodology to improve the temporal relation extraction between events. The work also proposes a guideline to annotate the causal relation between events. However, this approach requires dataset to be annotated with entities using their proposed annotation guideline, which is not applicable for social media short text such as tweets.

Recently, Kruengkrai et al. \cite{Kruengkrai2017} propose a method that utilizes background knowledge to determine causal relationship between two candidate events. The authors apply a \emph{multi-column neural network} \cite{Ciresan2012} to extract causal relationship between candidate phrases using archived web text. Word vectors of candidate phrases and background knowledge phrases are used as the features. However, this approach does not consider the spatial and temporal information of events. Another recent approach \cite{Rahimtoroghi2017} proposes to extract everyday events from user-generated text and identify causal relationship between events. The authors use a co-occurrence based technique to generate causal event pairs and calculate the causal strength of an event pair $(e_1, e_2)$ by calculating \emph{causal potential} (CP) as follows.
    
    \begin{equation}
    CP(e_1, e_2) = log\frac{P(e_2|e_1)}{P(e_1)} + log\frac{P(e_1\rightarrow e_2)}{P(e_2\rightarrow e_1)}
    \end{equation}
    
    Where $e_1$ and $e_2$ are two adjacent events. The adjacency is determined by applying 2-skip bigram model where two events occurring with two or less events are considered adjacent. A similar approach is used in \cite{Khan2017} to identify causal relationship between time series events extracted from computer event logs. The events have unique IDs and they may appear multiple times in the database. Authors exploit item set mining technique to detect pair of events with causal relationship. Then, the causal pairs with same effect event are merged together. Finally, the causal events for each merged relationships are sorted chronologically to generate the causal chain. One drawback of this approach is that it does not consider the causality between cause events while sorting the events for generating the causal chain. An event causality detection approach that is closely related to our approach is proposed by \cite{Ponti2017}, which uses a feed-forward neural network for detecting causality relation between events. The authors propose to enhance the feature set by calculating distances between event trigger word and other words in the sentence. However, for tweets, this positional information might not represent the causal strength properly as tweets often contain noisy characters and words e.g., emojis, hashtags and mentions and therefore, may not be applicable for event causality detection in tweets.

     \begin{figure*}
     \centering
      \begin{adjustbox}{minipage=0.99\textwidth,precode=\dbox}
        \includegraphics[scale=0.75]{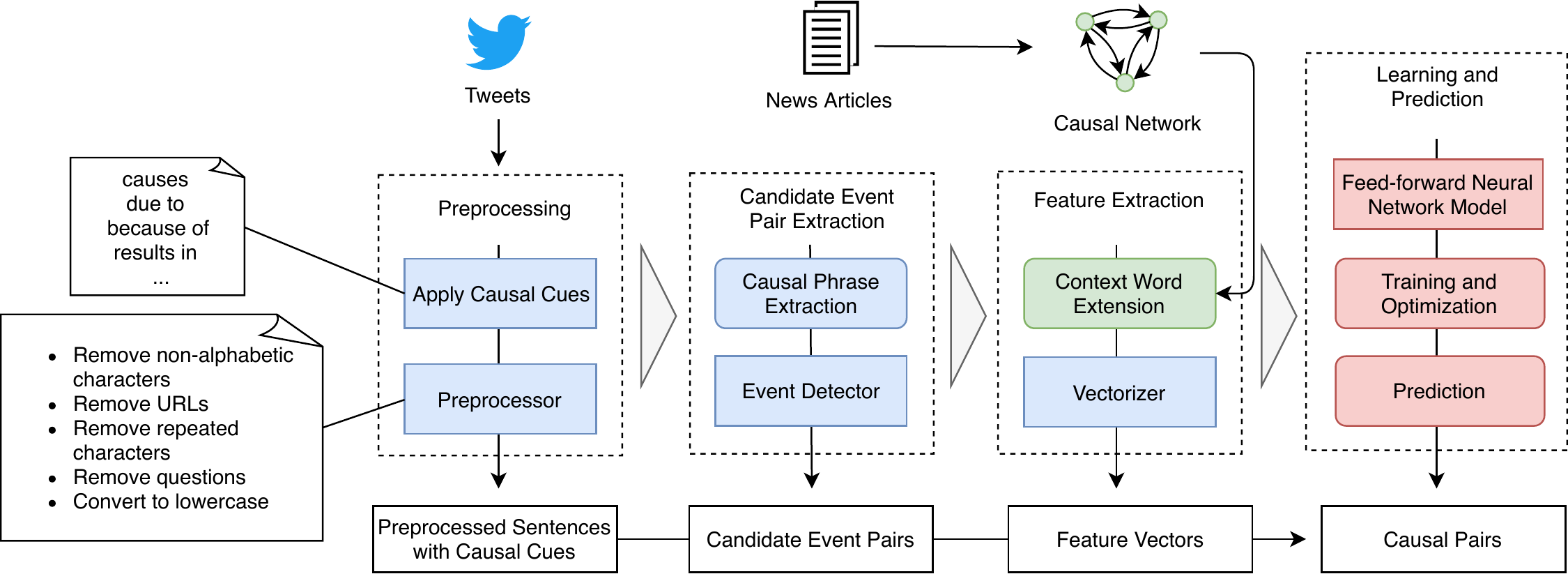}
        \end{adjustbox}
        \vspace{-1ex}
        \caption{An overview of the proposed method}
        \label{fig:system-overview}
        \vspace{-3ex}
    \end{figure*}

   \begin{table}[tb]
        \setlength{\tabcolsep}{1pt}
        \centering
        \caption{Representation of Events}
        \vspace{-2ex}
        \label{tab:event-representation}
        \begin{tabular}{|p{4.5cm}|p{4.2cm}|}
            \Xhline{3\arrayrulewidth}
            \multicolumn{1}{|c|}{\cellcolor{yellow}\textbf{Sentences}}  & \multicolumn{1}{c|}{\cellcolor{green}\textbf{Events}} \\ \hline \hline
            
            Storm \textbf{hits} Gold Coast  & hit (storm, coast, gold)     \\ \hline
            Mike \textbf{crashed} his car in Gold Coast  & crash (mike, car, coast, gold)       \\ \hline
            Heavy traffic \textbf{jam} in Gold Coast today        & jam (traffic, today, coast, gold)                   \\  \hline
            
            A \textbf{disruption} in bus service in Gold Coast due to \textbf{lack} of communication & 
            disruption (service, bus, coast, gold) \newline lack (communication, organizer, translink)\\
            
            \Xhline{3\arrayrulewidth}
        \end{tabular}
        \vspace{-3.5ex}
    \end{table}

\section{Problem Formulation}
    \label{sec:problem-formulation} 
    
    An event is a set of words that represents the occurrence of a specific incident. The event keyword is the word that actually triggers the event and the event attributes are the words that are syntactically related to the event keyword. Table \ref{tab:event-representation} displays a few example events\footnote{It should be noted that a sentence may have zero or more events.}. An event is considered as \emph{causal} if the event causes another event to occur. The other event is considered as the \emph{effect} of the causal event. For Example, ``\textit{\textbf{A disruption in bus service in Gold Coast} due to \textbf{lack of communication between translink and event organizers}}" contains a causal event and an effect event. In the above example the causal event is `lack of communication' and the effect event is `a disruption in bus service', where \emph{disruption} is the event keyword of the causal event and \emph{lack} is the event keyword of the effect event.  In this paper we focus on extracting such causal and effect event pairs from tweets and aim to address the following research question (RQ).

        \begin{enumerate}
        \setlength{\itemindent}{0.5cm}
        \item[\textbf{RQ:}] \textit{How to automatically detect causally related event pairs from social media short texts such as tweets?}
    \end{enumerate}

    Answering the above research question is however challenging as tweets are highly informal and prone to incorrect grammatical structure. For example, ``Much more worthwhile causes to use your time for" contains a causal cue word (`causes') but it is not expressing any event causality. Due to such challenges existing rule-based methods \cite{Luo2016, Sasaki2017} are less effective in event causality detection from short texts such as tweets. In addition, the events mentioned in tweets often lack information about the context e.g., ``India defeated Australia by 5 points" does not provide any information about the name of the sport the event is referring to. Hence, existing supervised learning based techniques such as the one proposed by Ponti et al. \cite{Ponti2017} have poor performances on tweets.

    We assume that causal and effect events occur in the same sentence in a tweet. We aim to detect such explicitly mentioned causal and effect event pairs. We define a candidate causal event as $e_1$ and effect event as $e_2$. Each event has a structure of $k(a_1, a_2, a_3, ..., a_n)$, where $k$ is the event keyword or trigger word and $\{a_i\}$ are the event attributes. For a candidate causal event pair $(e_1, e_2)$, our goal is to classify whether the event pairs have a causal relationship, i.e., $e_1$ causes $e_2$\footnote{Although $e_1$ causes $e_2$, it does not mean that $e_1$ is the only cause of $e_2$. There could be other causes of $e_2$, which is beyond the scope of this paper.}. Formally, we define the problem studied in this paper as follows:

    \begin{equation}
        \label{eq:causality-classification}
        f(e_1, e_2) = 
        \begin{cases}
            \text{Causal}  & \text{if } e_1 \text{ causes } e_2,\\
            \text {Not Causal} & \text{otherwise}
        \end{cases}
    \end{equation}

    where $f$ is a function that takes two events $e_1$ and $e_2$ as input and outputs either `Causal' or `Not Causal'. The function outputs `Causal' if $e_1$ causes $e_2$ in the input event pair and it outputs `Not Causal' otherwise.

\section{Our Approach}
    \label{sec:proposed-metnod}

    Our proposed method utilizes background knowledge to detect event causality. The background knowledge is extracted from news articles in the form of a causal network. To apply background knowledge, we extend event context words using the causal network. The events are then converted into word vectors to train a feed-forward neural network. The trained model is then used to detect causal relationship between a new pair of candidate causal events. Fig. \ref{fig:system-overview} illustrates an overall schematic overview of the proposed method.   
    
    \begin{table*}[tb]
    \setlength{\tabcolsep}{1.5pt}
    \small
        \centering
        \caption{Causal Cue Words}
        \vspace{-1ex}
        \label{tab:causal_cue_wordes}
        \begin{tabular*}{\textwidth}{@{\extracolsep{\fill}}|l|l|l|l|l|l|l|l|}
        \Xhline{2\arrayrulewidth}
        affect              & because    & causes       & due to                & if                & induce        & owing to       & results from  \\ \hline
        affected by         & because of & causing      & effect of             & if..., then       & induced       & reason for     & so that       \\ \hline
        affects             & bring on   & consequently & for this reason alone & in consequence of & inducing      & reason of      & that's why    \\ \hline
        and consequently    & brings on  & coz          & gave rise to          & in response to    & lead to       & reasons for    & the result is \\ \hline
        and hence           & brought on & coz of       & give rise to          & inasmuch as       & leading to    & reasons of     & thereby       \\ \hline
        as a consequence    & cause      & decrease     & given rise to         & increase          & leads to      & result from    & therefor      \\ \hline
        as a consequence of & caused     & decreased by & giving rise to        & increased by      & led to        & resulted from  & thus          \\ \hline
        as a result of      & caused by  & decreases    & hence                 & increases         & on account of & resulting from &               \\ \Xhline{2\arrayrulewidth}
        \end{tabular*}
        \normalsize
        \vspace{-3ex}
    \end{table*}

    \subsection{Tweet Preprocessing}
    \label{sec:preprocessing}

    As a first step of preprocessing, tweets are split into sentences. Sentences in tweets often contain characters that are considered as noise such as emojis, repeated characters and symbols. We perform a series of preprocessing steps to remove noisy characters from sentences. These steps include removal of non-alphabetic characters such as emojis, symbols, hashtags (`\#') and mention (`@') characters, and URLs. We also discard sentences ending with question mark (`?') and normalize repeated characters in a word, e.g., `yesss' to `yes'.

         \begin{figure*}[tb]
            
            \begin{adjustbox}{minipage=0.99\textwidth,precode=\dbox}
            \centering
            \includegraphics[scale=0.76]{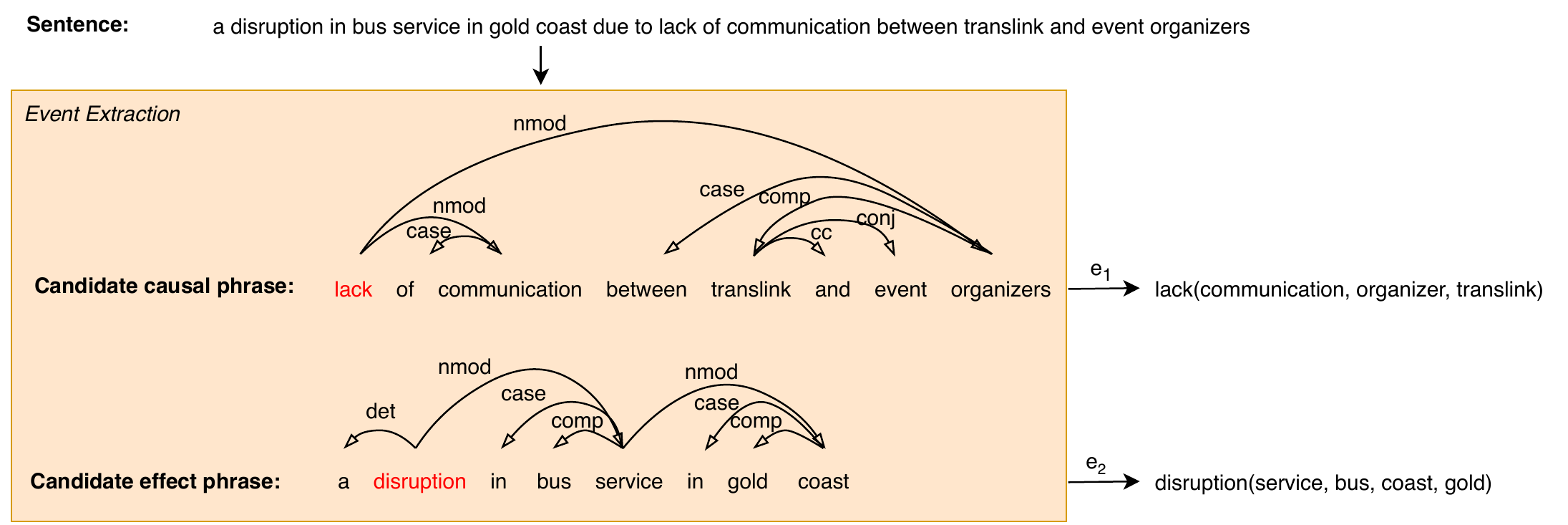}
            \end{adjustbox}
            \caption{An example of event pair extraction from a sentence}
            \label{fig:event-pair-extraction}
            \vspace{-2ex}
        \end{figure*}
        
        \begin{figure*}[tb]
            \begin{adjustbox}{minipage=0.99\textwidth,precode=\dbox}
            \centering
            \includegraphics[scale=0.62]{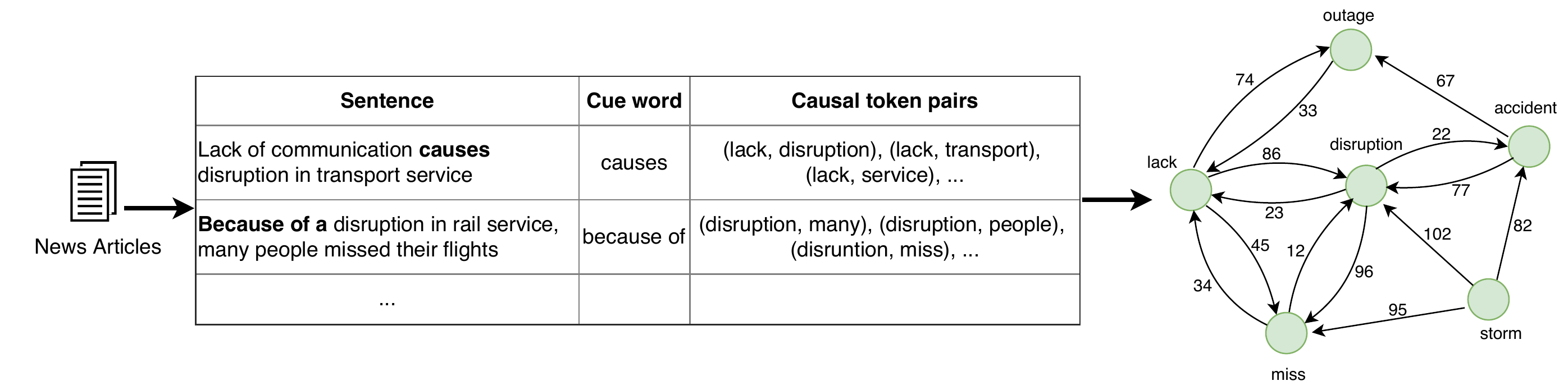}
            \end{adjustbox}
             \caption{Causal network construction from news articles}
            \label{fig:causal-netwrok}
            \vspace{-2ex}
        \end{figure*}

    \subsection{Event Pair Extraction}
        \label{sec:event-pair-extraction}

        In this step, a pair of candidate events is extracted from a sentence. At first, a sentence is split into candidate causal and effect phrases using a set of causal cue words \cite{Luo2016, Mirza2014} (please see Table \ref{tab:causal_cue_wordes}). For example, ``a disruption in bus service in gold coast due to lack of communication between translink and event organizers'' is split into ``a disruption in bus service in gold coast'' as the candidate effect phrase and ``lack of communication between translink and event organizers'' as the candidate causal phrase using the cue word \emph{due to}. Candidate cause and effect phrases are then  passed to the Stanford dependency parser \cite{chen2014} to detect the root word for each phrase. The root words are considered to be the event keyword $k$ of the corresponding events $e_1$ (or $e_2$). The other words that are related to the root word via `nsubj', `nsubjpass', `amod', `dobj', `advmod', `nmod', `xcomp', `compound:prt', `compound' and `neg' relationships are extracted as the event attributes $\{a_i\}$. We also extract other words that are related to the extracted event attributes via the above relationships as the event attributes. An example of event keyword $k$ and attributes $\{a_i\}$ extraction for the sentence ``\textit{\textbf{A disruption in bus service in Gold Coast} due to \textbf{lack of communication between translink and event organizers}}" is illustrated in Fig. \ref{fig:event-pair-extraction}.

           \subsection{Causal Network}
        \label{sec:causal-network}

       Background knowledge plays an important role in event causality detection. We use 1 million news articles\footnote{https://research.signalmedia.co/newsir16/signal-dataset.html} collected from the work of \cite{Signal1M2016} as a source of background knowledge and store the captured knowledge as causal relationships in a network called \emph{causal network}. To construct the network, first we extract the causal and effect phrases from article sentences using the causal cue words given in Table \ref{tab:causal_cue_wordes}. Then, the phrases are converted to lower cases after removing the stop words. The phrases are then tokenized and lemmatized. Each token in the either phrase represents a node in the network. A directed link from token A to token B contains frequency such that token A appeared in a causal phrase and token B appeared in the corresponding effect phrase, which is illustrated in Fig. \ref{fig:causal-netwrok}.

    \subsection{Context Word Extension}
        \label{sec:cotnext-word-extension}

\begin{figure*}[tb]
            \centering
            \includegraphics[scale=0.7]{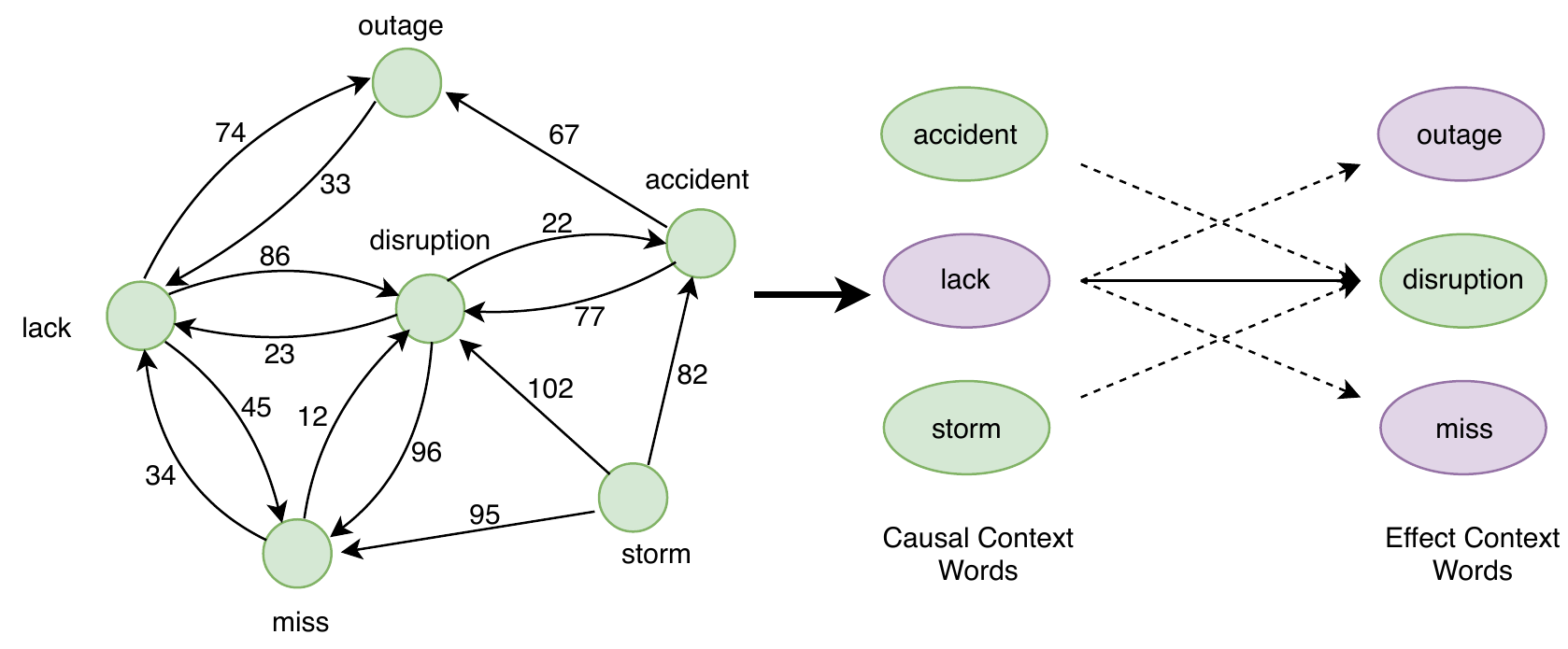}
            \vspace{-2ex}
            \caption{An example of $n$-word context word extension, where $n=2$ and the original candidate cause and effect keywords are \emph{lack} and \emph{disruption}, respectively}
            \label{fig:context-word-extension}
            \vspace{-3.5ex}
        \end{figure*}

        We use the background knowledge captured in the causal network to extend candidate event context words. For example, if there is a causal relationship between `Rain' and `Flood', it can be said that in many previous occasions the word \emph{Rain} was part of causal phrases where \emph{Flood} was a part of effect phrases. This knowledge can be applied to the causality detection method for extending context words. In our approach, which is pseudocoded in Algorithm \ref{alg:context-word-extension}, we look for the corresponding effect event keyword in the causal network to extend the context word $k$ of a candidate causal event. First, we identify a list of words with inward links to effect keyword in causal network. The list is then sorted in descending order of their frequencies. From that sorted list, we take top \emph{n} words to extend the context of candidate causal event $e_1$, where n is number of words we want to extend. Similarly, to extend the context of candidate effect event $e_2$, we identify the top \emph{n} effect words from causal network. A running example is given in Fig. \ref{fig:context-word-extension} to illustrate our context word extension technique for \emph{lack} and \emph{disruption} event keywords using the causal network. 
      
        \begin{algorithm}[t]
            \small
            \SetKwInOut{Input}{Input}
            \SetKwInOut{Output}{Output}
        	\SetKwInOut{Initialization}{Initialization}
            \Input{$e_1$: candidate causal event, $e_2$: candidate effect event, $n$: number of context word extension, $cnet$: causal network}
            \Output{$({e_1}', {e_2}')$: list of events with expanded context words}
            
            $k_1 \gets get\_event\_keyword(e_1)$\;
            $k_2 \gets get\_event\_keyword(e_2)$\;
            $ct \gets get\_causal\_terms(cnet, k_2)$\;
            $et \gets get\_effect\_terms(cnet, k_1)$\;
            ${e_1}' \gets list(k_1)$ \tcp*[r]{create a list with $k_1$}
            ${e_2}' \gets list(k_2)$ \tcp*[r]{create a list with $k_2$}
            \For{$i \gets 0$ to $n-1$} {              
                ${e_1}' \gets {e_1}' + list(ct[i])$ \tcp*[r]{append terms}
                ${e_2}' \gets {e_2}' + list(et[i])$ \tcp*[r]{append terms}
            }
            
            ${e_1}' \gets {e_1}' + list(get\_attributes(e_1))$\;
            ${e_2}' \gets {e_2}' + list(get\_attributes(e_2))$\;

            return $({e_1}', {e_2}')$ \tcp*[r]{events with expanded context words}
            \caption{Context Word Extension}
            \label{alg:context-word-extension}
            \normalsize
        \end{algorithm}

    \subsection{Feature Extraction}
    \label{sec:feateure-extraction}

    In the feature extraction stage, the candidate causal event $e_1$ and the candidate effect event $e_2$ are converted into a numerical vector $v$. However, before the conversion, the context words of $e_1$ and $e_2$ are extended following the steps described in Section \ref{sec:cotnext-word-extension}, which generates $e_1'$ and $e_2'$ respectively. To convert $e_1'$ and $e_2'$ into $v$, we train a Word2vec model \cite{mikolov2013} from 1 million news articles (the same dataset that is used to build the \emph{causal network}). Then we extract the dictionary of words $\mathcal{D}$ from the trained Word2vec model. Using this dictionary we replace every word in $e_1'$ and $e_2'$ by its corresponding index in $\mathcal{D}$. The word indexes of $e_1'$ and $e_2'$ are then concatenated together to construct a single index vector $i_v$. In the next step, each index $i_v$ is replaced by its corresponding word embedding which produces a matrix of word embeddings $M$. The number of columns in $M$ is 300 and the number of rows is the same as the total words $e_1'$ and $e_2'$. Finally, the matrix $M$ is flattened by taking mean and converted to a vector $v$ of size 300. This vector $v$ is passed to the input layer of the feed-forward neural network for training and detection, which is discussed below.    
    
     \begin{figure*}[tb]
        \vspace{-2ex}
            \begin{adjustbox}{minipage=0.98\textwidth,precode=\dbox}
            \centering
            \includegraphics[scale=0.76]{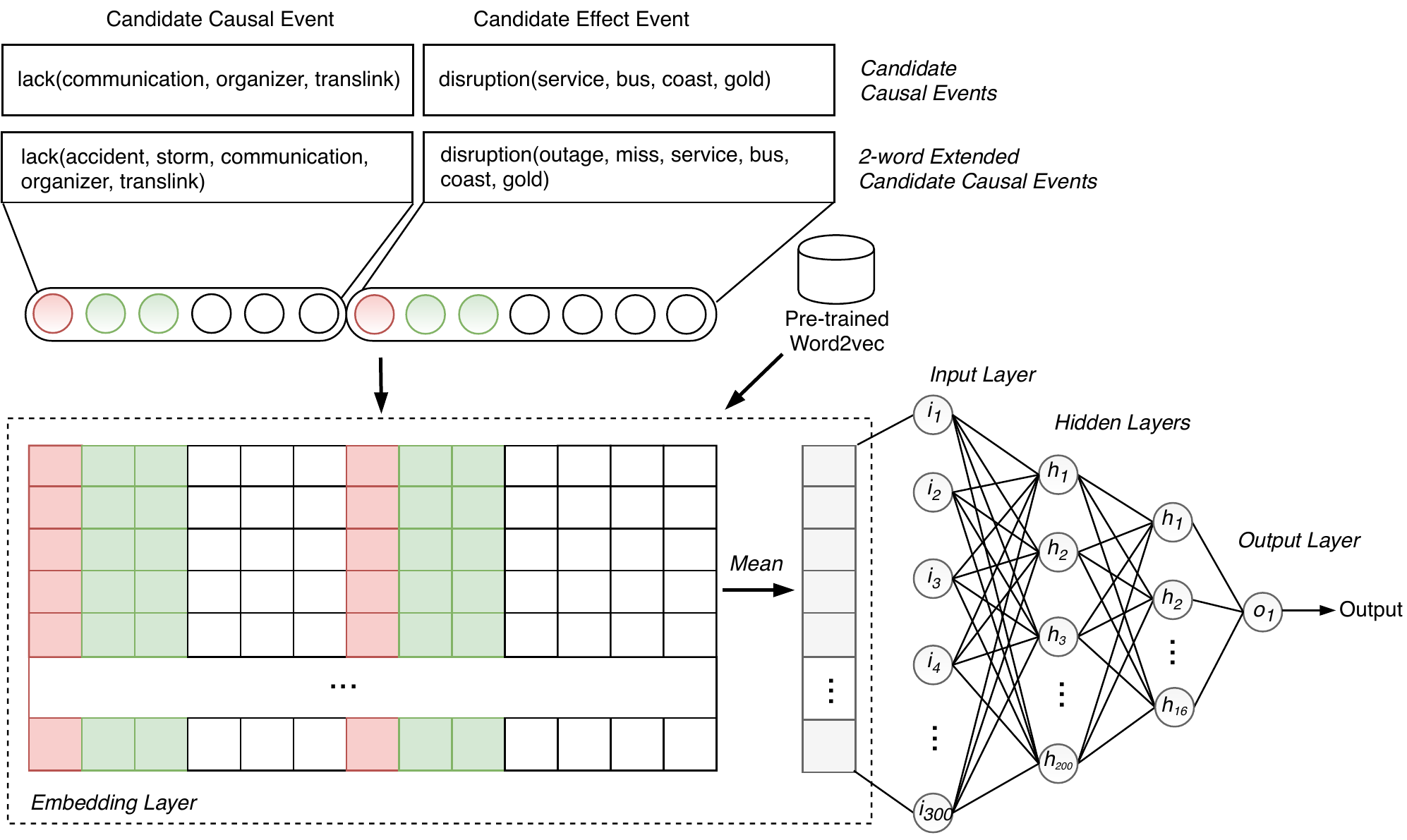}
            \end{adjustbox}
            \caption{Feature extraction from an event pair and different layers of the feed-forward neural network}
            \label{fig:neural-network-training}
            \vspace{-4ex}
    \end{figure*}

    \subsection{Learning and Detection}

    \textbf{Learning the Detection Model}. In this step, we train a feed-forward neural network model. First, we prepare a gold standard dataset that contains event pairs where each pair is labeled as either `causal' or `not causal'. Then we extract feature for the candidate event pair following the steps described in \ref{sec:feateure-extraction}, which includes context word extension and vectorization. In context word extension step, we extend the event context word $k$ for both causal event $e_1$ and effect event $e_2$ using a pre-constructed causal network (please see Sec. \ref{sec:cotnext-word-extension}). The context word extension step generates $e_1'$ and $e_2'$ where $e_1'$ is the extended version of $e_1$ and $e_2'$ is the extended version of $e_2$. After performing the context word extension, every event pair is converted into a 300 dimensional feature vector following the steps described in Sec. \ref{sec:feateure-extraction}. Such feature vectors of all candidate event pairs and their corresponding labels (`causal' or `not causal') are passed to a feed-forward neural network for training. The trained model is then used to detect the causal relationship between the candidate event pairs in unknown tweets.

    \textbf{Causal Relationship Detection}. The causal relationship detection between event pairs in an unknown tweet starts with a series of preprocessing steps as described in Section \ref{sec:preprocessing}. After preprocessing, the tweet is converted into a set of sentences where noisy characters such as emojis, hashtags (`\#') and mentions (`@') are removed. We also remove question sentences with the assumption that questions do not contains an event causal relationship. The sentences are then passed to the candidate event pair extraction steps where pairs of candidate causal events are extracted (please see Section \ref{sec:event-pair-extraction}). The next step is to extract features, where context word extension technique is applied to both candidate causal event and effect event (please see Section \ref{sec:cotnext-word-extension}). The event words are then converted into the feature vector (please see Section \ref{sec:feateure-extraction}) which is passed to the trained feed-forward neural network model for event causality detection. 
    
    The schematic diagram of causal relationship detection in candidate event pairs, which includes learning the neural network model as well as the detection of causal relationship in unknown event pairs, is  illustrated in Fig. \ref{fig:neural-network-training}.

    \section{Experiments}
    \label{sec:experiments}
    This section presents our experiments and demonstrates the effectiveness of our method on event causality detection.
    
\begin{figure*}[tb]
\centering
\setlength{\tabcolsep}{2pt}
\begin{tabular}{ccc}
\begin{minipage}[a]{0.32\linewidth}
\centering
\subfloat[0-word]{\includegraphics[scale=0.38]{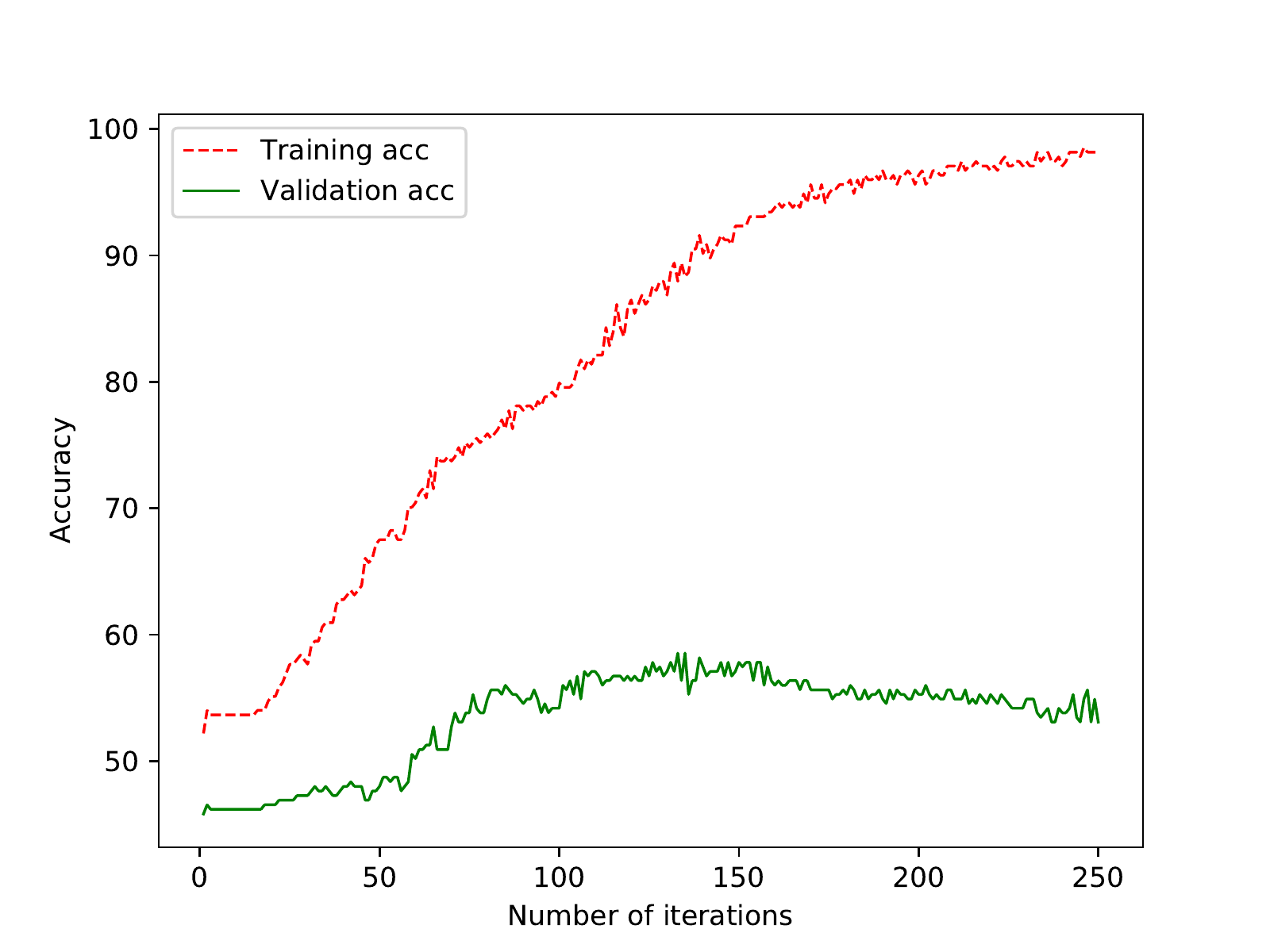}}
\end{minipage}
&
\begin{minipage}[a]{0.33\linewidth}
\centering
\subfloat[1-word]{\includegraphics[scale=0.38]{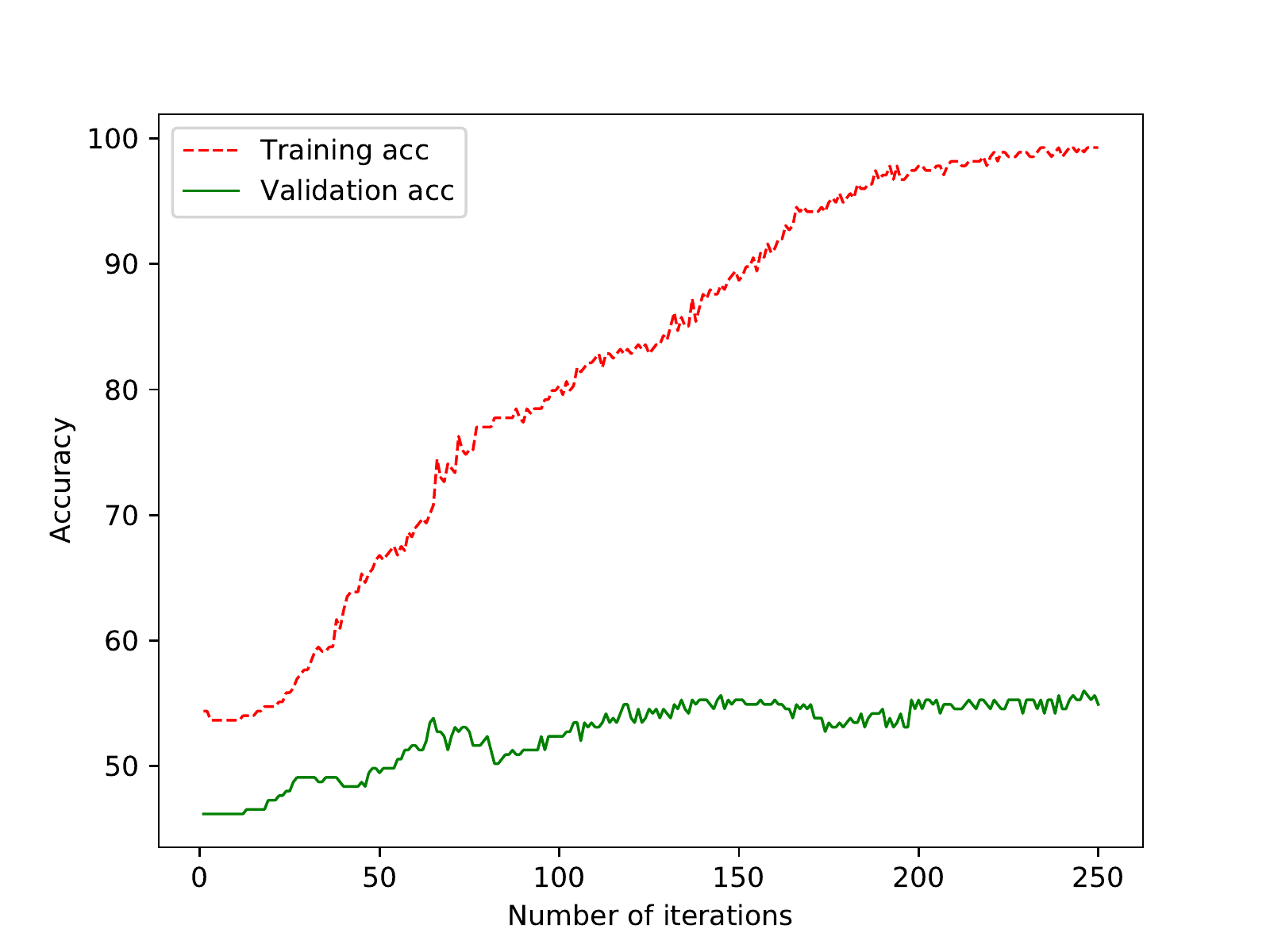}}
\end{minipage}
&
\begin{minipage}[a]{0.32\linewidth}
\centering
\subfloat[2-word]{\includegraphics[scale=0.38]{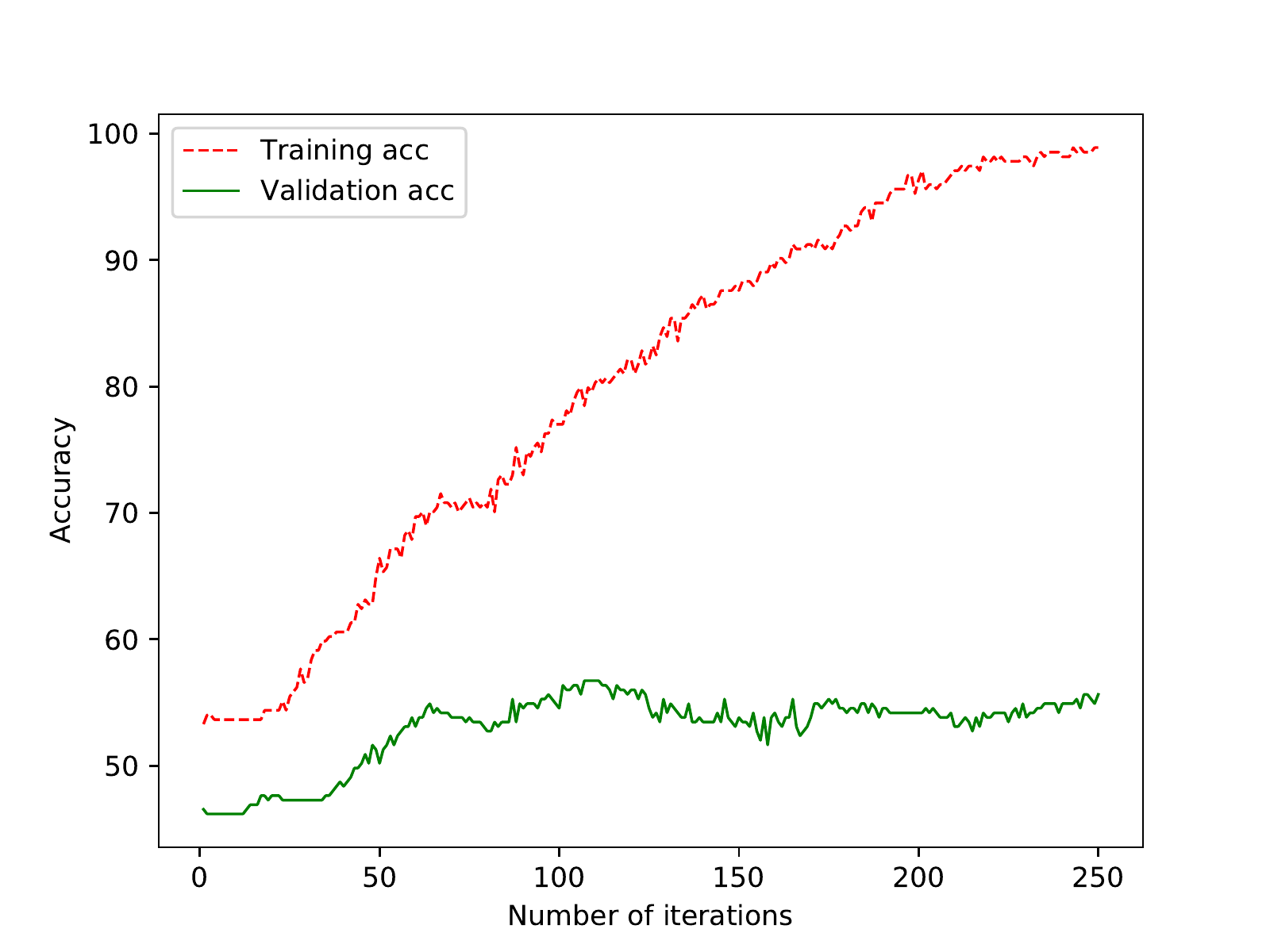}}
\end{minipage}\\
\begin{minipage}[a]{0.32\linewidth}
\centering
\subfloat[3-word]{\includegraphics[scale=0.38]{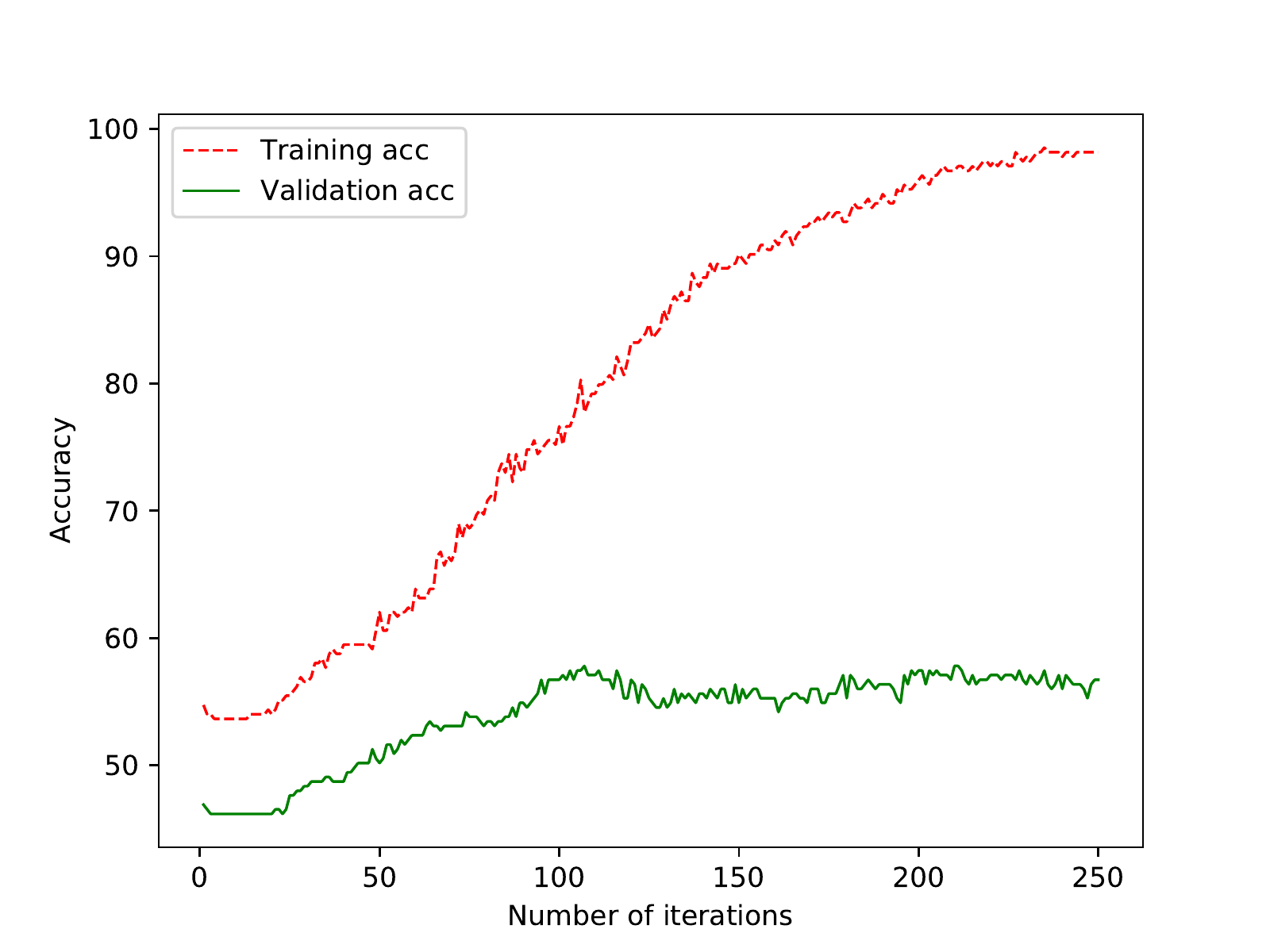}}
\end{minipage}
&
\begin{minipage}[a]{0.32\linewidth}
\centering
\subfloat[4-word]{\includegraphics[scale=0.38]{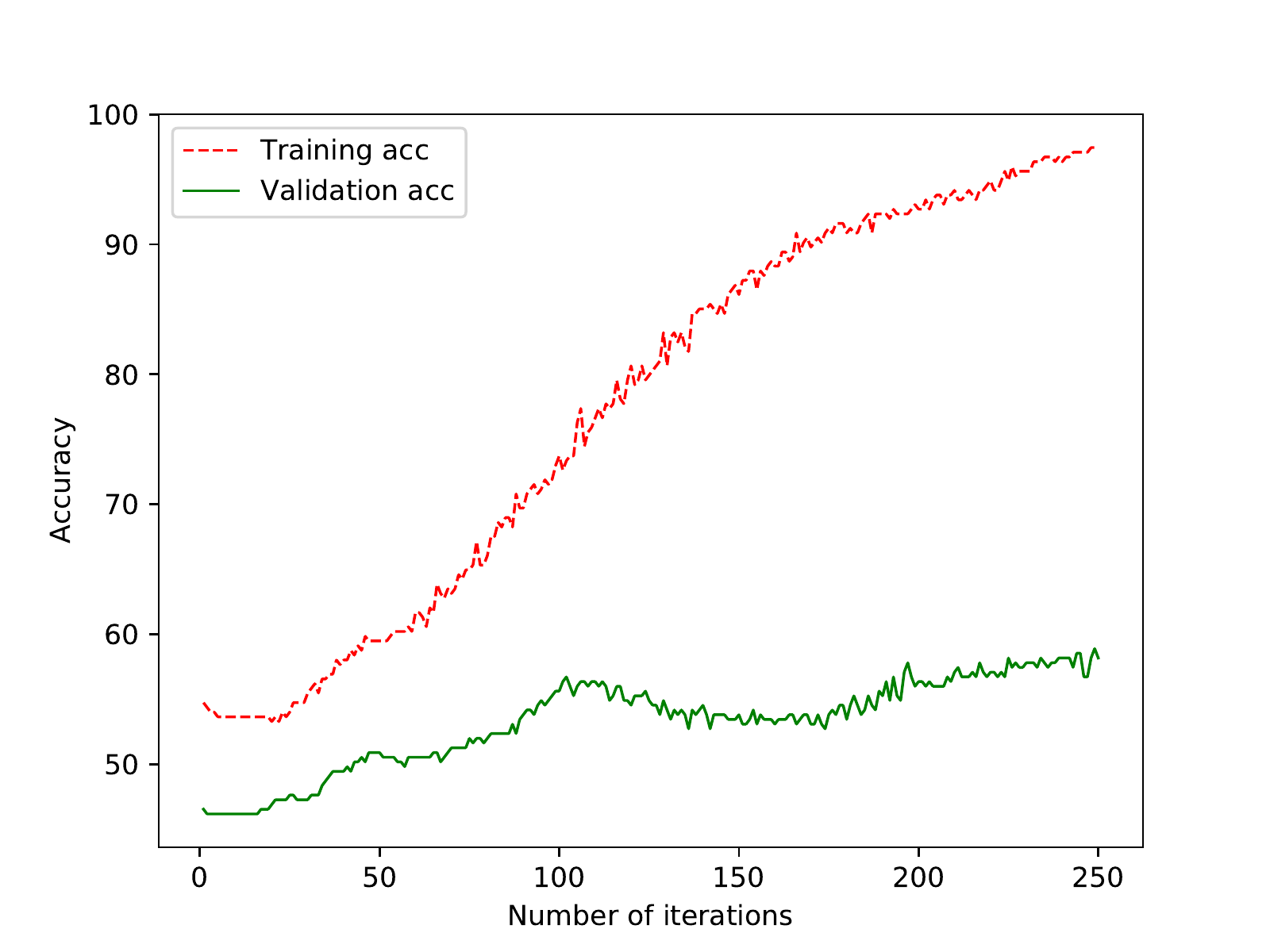}}
\end{minipage}
&
\begin{minipage}[a]{0.32\linewidth}
\centering
\subfloat[5-word]{\includegraphics[scale=0.38]{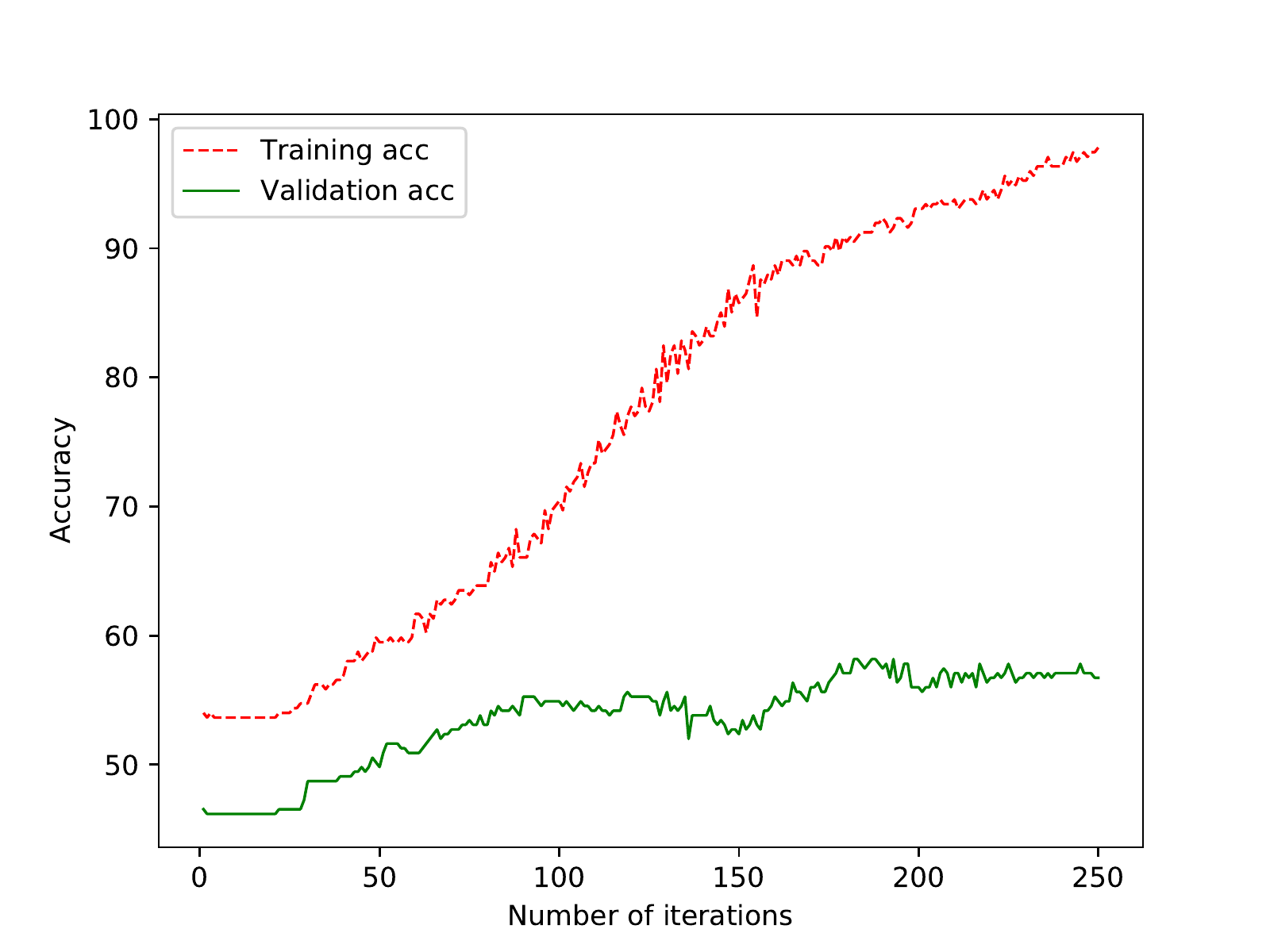}}
\end{minipage}
\end{tabular}
\vspace{-1ex}
\caption{Training accuracy optimization for models from 0 to 5-word extension}
\label{fig:model-accuracy-optimization}
\vspace{-3.5ex}
\end{figure*}
    
    \subsection{Dataset}
    We collect 207,705 tweets that are related to the Commonwealth Games 2018 in Gold Coast, Australia and posted during the period from 2017-10-05 to 2018-05-07 using twitter API\footnote{https://developer.twitter.com/en/docs/tweets/search/overview}. The following hashtags are used as keywords to collect the tweets: `\#CommonwealthGames', `\#CommonwealthGames2018', `\#GC2018', and  `\#ShareTheDream'. After performing the preprocessing steps mentioned in Section \ref{sec:preprocessing}, we identify 913 candidate cause and effect event pairs based on the approach described in Section \ref{sec:event-pair-extraction} and annotate them manually as either `Causal' or `Not Causal'. 
    
    We split our annotated dataset to separate 60\% data for training and 40\% for testing. We ensure that the ratio of `Causal' and `Not Causal' data remains same in both training and test data. The statistics of the tested dataset is presented in Table \ref{tab:dataset-statistics}. Among the training data (60\% of the original dataset), we use 50\% data for learning the model and the rest 50\% data for validation and parameter optimization.
    
    
    \begin{table}[t]
    \centering
    \caption{Statistics Of The Dataset}
    \label{tab:dataset-statistics}
    \begin{tabular}{|l|c|c|}
    \Xhline{2\arrayrulewidth}
    \textbf{Set} & \textbf{Causal} & \textbf{Not causal} \\ \hline \hline
    Full dataset & 459  & 457 \\ \hline
    Training   & 275  & 274  \\ \hline
    Test      & 184  & 183   \\ \hline
    \Xhline{2\arrayrulewidth}
    \end{tabular}
    \vspace{-4ex}
    \end{table}

\subsection{Setup}
We implement the proposed method in Python 3.6 and use keras python package\footnote{https://keras.io/} to implement the feed-forward neural network based causality detection method. The neural network has an input layer, two hidden layers and an output layer. The input layer contains 300 nodes and we use `ReLU' as the activation function, which accepts 300 dimensional event vectors as the input. Next to this layer we have two fully connected hidden layers that consists of 200 nodes and 16 nodes, respectively (this topology has been optimized empirically). Each node in both of the hidden layers uses `ReLU' as the activation function. The output layer is another fully connected layer that contains only one node and `sigmoid' function as the activation function. Empirically, we use ADADELTA \cite{zeiler2012} algorithm as the cost function optimizer, batch size 40. We use the evaluation metric: 'Accuracy', to optimize the parameters.

\subsection{Parameter Optimization}
\label{sec:parameter_optimization}
We optimize the learning rate of our   feed-forward neural network model while keeping other parameters fixed. Initially, we set our learning rate to $0.01$ and raise the learning rate gradually if the model learns too much detail and overfit the training data. We find the best learning performance for learning rate $0.1$. We perform this experiment for $0$ to $5$ event context word extensions and run for 250 iterations. The model with 0-word extension corresponds to the model where no event context word extension is applied. Fig. \ref{fig:model-accuracy-optimization} illustrates training and validation accuracy for different number of iterations. We observe that the validation accuracy stops growing or starts to decline in between 100 to 200 iterations. As the validation accuracy of the model does not improve for more iterations and model only overfits the training data, we choose to stop training at 150 iterations.

\begin{figure}[tb]
\centering
\includegraphics[scale=0.53]{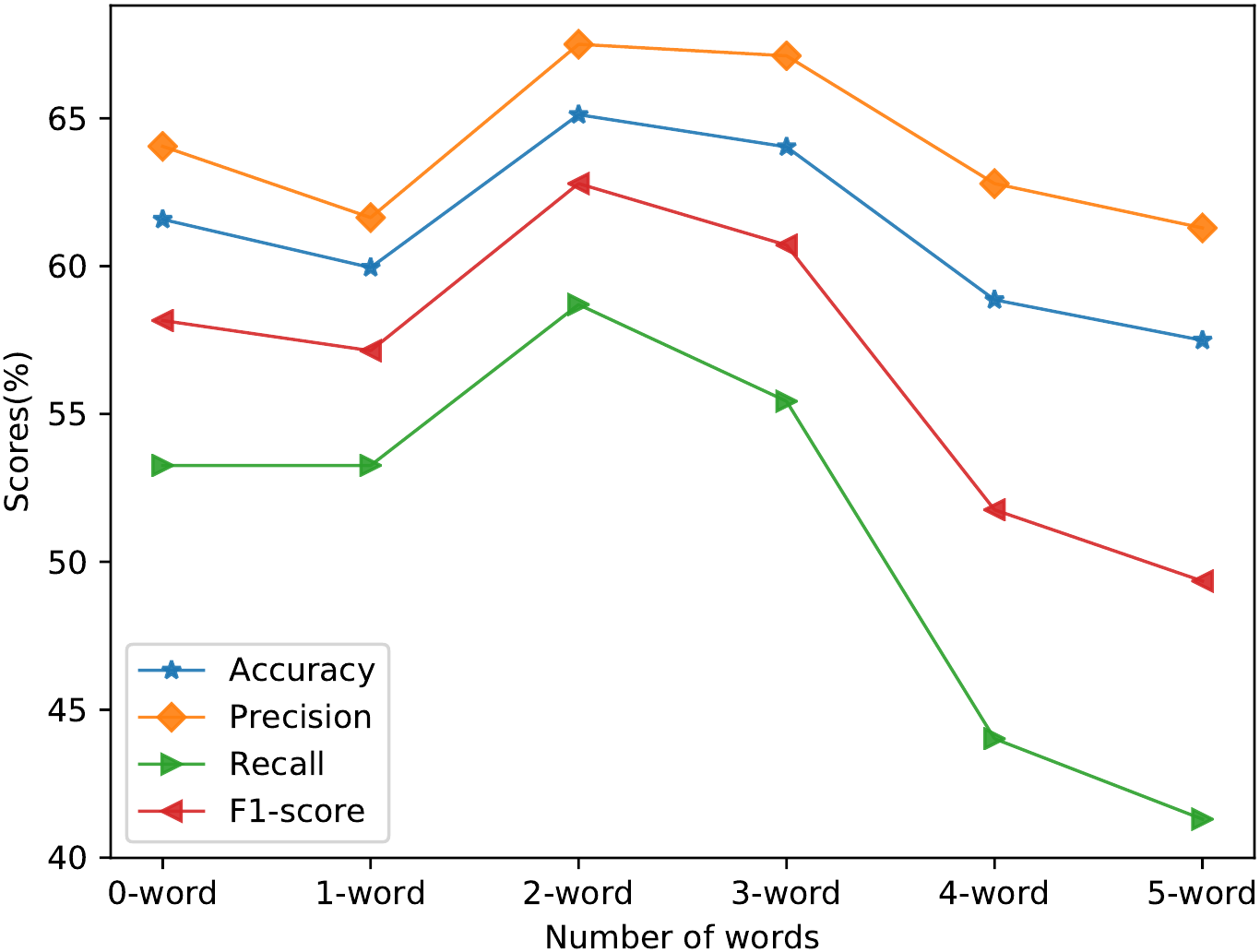}
\vspace{-1ex}
\caption{\label{fig:scores-word-extension-news}Evaluation scores of different settings in the proposed event context word extension technique}
\vspace{-4.5ex}
\end{figure}

\subsection{Performance Evaluation}
\begin{figure}[tb]
\centering
\includegraphics[scale=0.53]{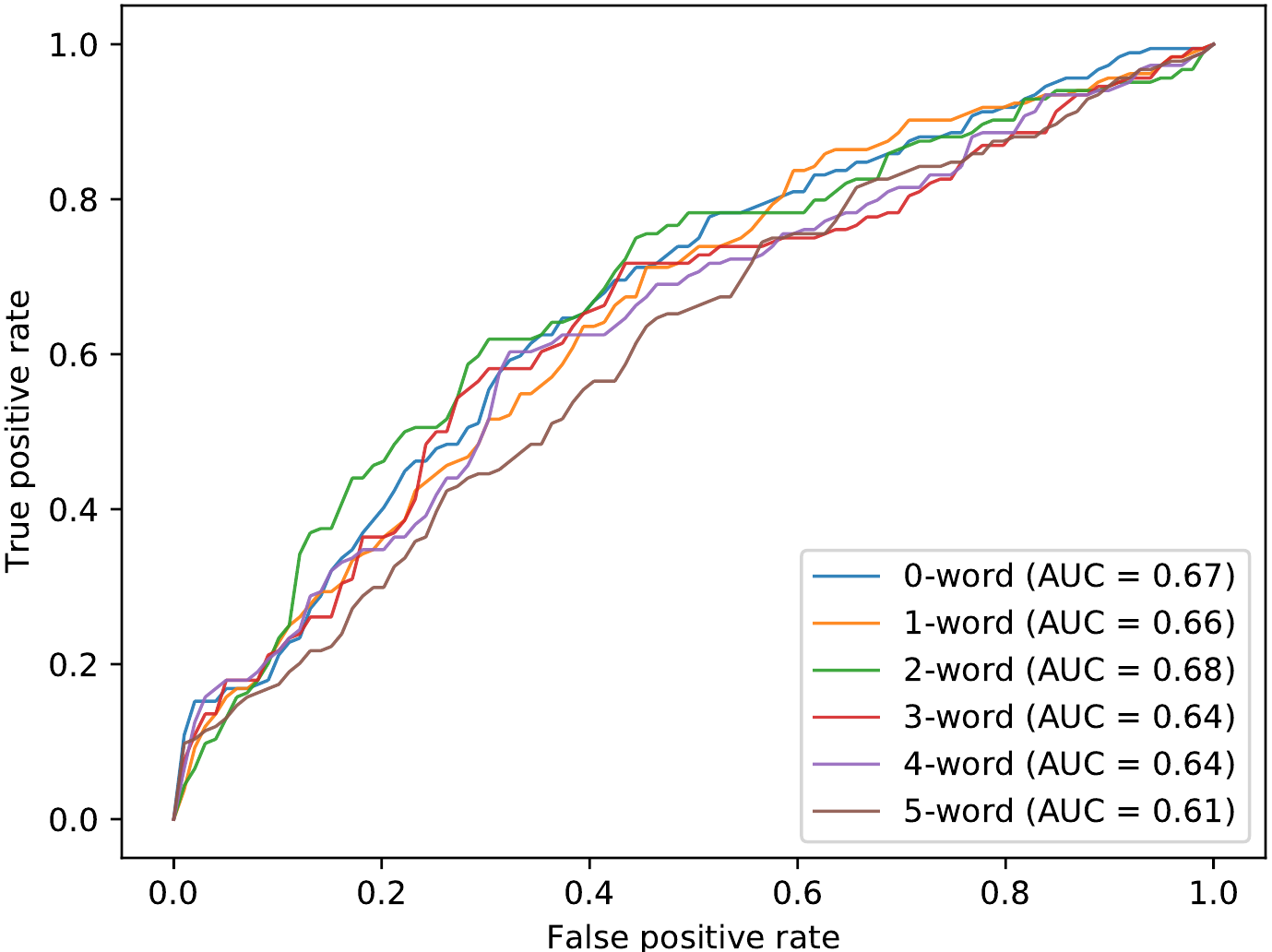}
\vspace{-1ex}
\caption{ROC Curve of different settings in the proposed event context word extension technique}
\label{fig:roc-curve-word-extension-news}
\vspace{-1ex}
\end{figure}

To evaluate the generalizability of our proposed method of event causality detection on unseen tweets, first we train the proposed neural network model on the training set (60\%) using the optimized parameters learned during the training phase (as explained in Section \ref{sec:parameter_optimization}). Then, we compare the performance of the proposed method for event context word extension against the method that uses feature vectors with no context word extension for the test dataset (40\%). Fig. \ref{fig:scores-word-extension-news} illustrates the standard evaluation scores: accuracy, precision, recall and F1-score of different settings of the event context word extension. The results suggest that we gain performance improvement across the evaluation scores for both 2 and 3-word extension compared to the model that uses 0-word extension. The model with 2-word extension achieves the best evaluation scores. We also observe that increasing the number of word extension after 3 does not perform well and the performance drops sharply. This is because the words extracted from the knowledge base become more prominent than the original event words. We also generate an ROC curve to compare the performance of our proposed method against the 0-word extension based method as given in Fig. \ref{fig:roc-curve-word-extension-news}. From Fig. \ref{fig:roc-curve-word-extension-news}, we see that the Area Under Curve (AUC) value for 2-word extension based method is higher than the 0-word extension based method. In conclusion, the 2-word extension based method is the best performer. 

\begin{table}[t]
\centering
\small
 \setlength{\tabcolsep}{.23pt}
\caption{Comparison of the proposed method with existing approaches}
\vspace{-1ex}
\label{tab:comparison-with-benchmarks}
\begin{tabular}{|l|c|c|c|c|}
\Xhline{2\arrayrulewidth}
\multicolumn{1}{|c|}{\textbf{Methods}} & \multicolumn{1}{c|}{\textbf{Accuracy}} & \multicolumn{1}{c|}{\textbf{Precision}} & \multicolumn{1}{c|}{\textbf{Recall}} & \multicolumn{1}{c|}{\textbf{F1-score}} \\ \hline \hline
Commonsense \cite{Luo2016}                            & 50.95                                  & 56.67                                   & 9.24                                & 15.89                                  \\ \hline
Commonsense + Multi-word \cite{Sasaki2017}                            & 50.14                                  & 54.55                                   & 3.26                                & 6.15                                  \\ \hline
FFNN + Position \cite{Ponti2017}                            & 50.00                                  & 52.38                                   & 6.08                                & 10.89                                  \\ \hline
FFNN + 2-word Extension (ours)                                    & \textbf{65.94}                                  & \textbf{67.46}                                 & \textbf{61.96}                                & \textbf{64.59}                                  \\ \hline
\Xhline{2\arrayrulewidth}
\end{tabular}
\vspace{-3.5ex}
\end{table}

To compare the performance of the proposed method with other existing methods, we implement three existing causality detection systems. We implement the commonsense causality detection method (Commonsense) proposed by Luo et al. \cite{Luo2016}. The method takes a set of candidate causal phrases as input and decides if there is a causal relationship between each pair of phrases by calculating their causal strength based on the knowledge available in the causal network. For this approach, we use the same causal network that we use for our proposed method. Additionally, we implement another approach proposed by Sasaki et al. \cite{Sasaki2017} that extends the commonsense causality detection method for multi-word phrases (Commonsense + Multi-word). Our third existing method is an event causality detection system \cite{Ponti2017} that enhances the feature set of a feed-forward neural network using the positions of words in the sentence (FFNN + Position). We set iterations to 150, learning rate to 0.1 and batch size  to 10 to train the neural network for this method. The experimental results are given in Table \ref{tab:comparison-with-benchmarks}. We observe better performance of the proposed system compared to the existing state-of-the-art event causality detection systems such as Commonsense  \cite{Luo2016}, Commonsense+Multi-word \cite{Sasaki2017} and FFNN+Position \cite{Ponti2017} based systems. From Table \ref{tab:comparison-with-benchmarks}, it is evident that the performance gain achieved by our method is at least 570\% and 306\% in terms of Recall and F1-score, respectively. This outcomes demonstrate that the proposed event context word extension technique is capable of overcoming the issue of insufficient context information in candidate causal event pairs in tweets.

 \section{Conclusion}
 \label{sec:conclusion}
This paper proposes a feature enhancement technique for supervised learning based event causality detection approach. We demonstrate that commonsense background knowledge can be used to extend event context information, which helps to enhance feature set of a supervised learning based method. We develop a neural network based event causality detection method that uses event context word extension technique to detect causality relationship between pair of events. We find that the neural network based model performs better when trained on the enhanced feature set for event causality detection in tweets. In this paper, we focus on extracting pairs of causally related events from tweets, but users often post opinionated or sarcastic posts which may contain incorrect event causality relationships. We aim to perform fact checking of event causality relationships in tweets as future works. 

 \bibliography{library} 
 \bibliographystyle{ieeetr}
\end{document}